\documentclass{article}
\usepackage[english]{babel}
\usepackage{amsmath,amsfonts,amsthm,amssymb,amscd}
 \usepackage{verbatim}

\renewcommand{\Re}{\mathop{\rm Re}\nolimits}
\renewcommand{\Im}{\mathop{\rm Im}\nolimits}

\newcommand{\e}{\varepsilon}

\newcommand{\Q}{{\mathbb Q}}
\newcommand{\C}{{\mathbb C}}
\newcommand{\R}{{\mathbb R}}

\newcommand{\pP}{{\mathbb P}}

\newcommand{\E}{{\mathbb E}}

\newcommand{\la}{\lambda}

\newcommand{\ty}{\infty}

\newcommand{\aA}{{\cal A}}

\newcommand{\DD}{{\cal D}}

\newcommand{\FF}{{\cal F}}

\newcommand{\KK}{{\cal K}}
\newcommand{\LL}{{\cal L}}

\newcommand{\PP}{{\cal P}}
\newcommand{\QQ}{{\cal Q}}
\newcommand{\RR}{{\cal R}}

\newcommand{\UU}{{\cal U}}
\newcommand{\VV}{{\cal V}}

\newcommand{\lag}{\langle}
\newcommand{\rag}{\rangle}

\newcommand{\dd}{{\textup d}}

\newcommand{\PPPP}{{\mathfrak P}}

\newcommand{\BBBBB}{{\mathcal B}}

\newcommand{\supp}{\mathop{\rm supp}\nolimits}

\theoremstyle{plain}
\newtheorem{theorem}{Theorem}[section]
\newtheorem{lemma}[theorem]{Lemma}
\newtheorem{proposition}[theorem]{Proposition}

\newtheorem{condition}[theorem]{Condition}
\theoremstyle{remark}
\newtheorem{remark}[theorem]{Remark}

\newcommand{\de}{\delta}
\numberwithin{equation}{section}
\begin{document}
 \author{Vahagn Nersesyan}
\date{}
\title{Global approximate controllability for Schr\"odinger equation in higher Sobolev norms and applications}
 \maketitle

\begin{center}
 Laboratoire de Math\'ematiques,
Universit\'e de Paris-Sud XI\\ B\^atiment 425, 91405 Orsay Cedex,
France\\ E-mail: Vahagn.Nersesyan@math.u-psud.fr
\end{center}
 {\small\textbf{Abstract.} We prove that the  Schr\"odinger equation is approximately controllable in Sobolev spaces $H^s$,
  $s>0$ generically with respect to the potential. We give two applications of this result.
   First, in the case of one space dimension, combining our result with a local exact controllability property, we get the global exact controllability of the system in higher Sobolev spaces.
 Then we prove that the  Schr\"odinger equation with a potential which has a   random time-dependent amplitude
 admits at most one stationary measure on   the unit
sphere $S$ in $L^2 $.
  }\\\\
  \tableofcontents

\section{Introduction}\label{S:intr}
In this paper, we study the problem
\begin{align}
i\dot z   &= -\Delta z+V(x)z+ u(t) Q(x)z,\,\,\,\,x\in D,\label{E:hav1}\\
z\arrowvert_{\partial D}&=0,\label{E:ep1}\\
 z(0,x)&=z_0(x),\label{E:sp1}
\end{align} where $D\subset\R^m$ is a bounded domain with smooth boundary, $V \in C^\ty(\overline{D},\R)$ is an arbitrary   given function,  $u$ is the control,
and $z$ is the state. We prove that this system is approximately
controllable to the eigenfunctions of $-\Delta+V$ in   Sobolev
spaces $H^s$, $s>0$ generically with respect to the potential $Q$.
In the case $m=1$ and $V=0$, combination of our result with the
local exact controllability property obtained by Beauchard
\cite{BCH} gives the global exact controllability of the system in
the spaces $H^{5+\e}$. Approximate controllability property
implies also that the random  Schr\"odinger equation
  admits at most one stationary measure on the    unit
sphere $S$ in $L^2 $.

The problem of   controllability of the Schr\"odinger equation has
been largely studied in the literature.
 Let us mention some previous   results closely related to the present paper.
  Ball,  Marsden and Slemrod
\cite{BMS} show  that the set of attainable points from any
initial data in $S\cap H^2$ by system (\ref{E:hav1}),
(\ref{E:ep1})  admits a dense complement in $S\cap H^2$.
 In \cite{BCH}, Beauchard  proves an
exact controllability result for the system with  $m=1,V(x)=0$ and
$Q(x)=x$
   in $H^7$-neighborhoods of the  eigenstates.  Beauchard and Coron
\cite{BeCo} established later a partial global exact
controllability result, showing that the system in question
  is also controlled between some
neighborhoods of any two eigenstates. Chambrion et al. \cite{CH}
and Privat and Sigalotti \cite{PISI} prove that (\ref{E:hav1}),
(\ref{E:ep1})  is  approximately controllable in~$L^2$ generically
with respect to the potentials    $V,Q$ and the domain $D$.   See
also the papers  \cite{RSDR,TR,ALT,ALAL,AC,BCMR}   for
controllability of finite-dimensional systems and the papers
\cite{L,MZ,BP,Z,DGL,Mir,BM,PERV} for controllability properties of
various  Schr\"odinger systems.

 Let us recall that, in the case of the space $H^2$, we established a stabilization property for system
 (\ref{E:hav1}), (\ref{E:ep1}) in \cite{VN}. Namely,  we introduce a  Lyapunov function $\VV(z)\ge0$ that controls the
$H^2$-norm of $z$ and possesses the following properties:
\begin{enumerate}
\item[$\bullet$] $\VV(z)=0$ if and only if $z=ce_{1,V}$, where  $e_{1,V}$ is the first eigenfunction of the operator
$-\Delta+V$,
\item[$\bullet$] $\frac{\dd }{\dd t}\VV(z(t))=uG(z(t))$, where $z(t)=z(t,u)$ is the solution of (\ref{E:hav1})-(\ref{E:sp1}) and $G(z)$ is
a function given explicitly.
\end{enumerate}Choosing the feedback low $u(z)=-G(z)$, we see that $\frac{\dd }{\dd t}\VV(z(t))=-u^2\le~0$, i.e., the function
$\VV$ decreases on the trajectories of (\ref{E:hav1})
corresponding to the feedback $u$. Thus, to conclude, it
suffices to prove that $\VV(z(t))\rightarrow0$. To this end, we
use an iteration argument and show that the $H^2$-weak
$\omega$-limit set of any trajectory $z(t)$ contains a minimum
point of the function $\VV$, i.e., the eigenfunction $ce_{1,V}$,
where $c\in\C, |c|=1$. Thus we construct explicitly a feedback low
$u(z)$ which forces the trajectories
 of the system to converge to the eigenstate $\{ce_{1,V}:c\in\C,|c|=1\}$.

The aim of this paper is to generalize these ideas to the case of
the spaces $H^k,$ $k>2$. The main difficulty is  that we are not
able to
 construct a  Lyapunov function $\VV(z)$ such that $\frac{\dd }{\dd t}\VV(z(t))=uG(z(t))$ for some function $G$.
 However,   notice that  for any $w\in C^\ty([0,T],\R)$ we can calculate explicitly the derivative
 $ \frac{\dd}{\dd \sigma}\VV(z(t,\sigma w))|_{\sigma=0}$. We show that there is a time $T>0$ and a control $w$ such that
  $\frac{\dd}{\dd \sigma}\VV(z(t,\sigma w))|_{\sigma=0}\neq0$. Hence we can choose $\sigma_0$
  close to zero such that
  $$
\VV(z(T,\sigma_0 w))<\VV(z(T,0))=\VV( z_0 ).
$$Thus for any point $z_0$ we find a time $T>0$ and a control $u$ such that $$\VV(z(T,u))<\VV( z_0 ).$$
Using an  iteration argument close to that of \cite{VN}, we
conclude that there are sequences $T_n>0$ and $u_n\in
C^\ty([0,T_n],\R)$ such that $z(t_n,u_n)\rightarrow e_{1,V}$. Thus
any point $z_0$ can be approximately controlled to $e_{1,V}$.

Then, in the case $m=1$ and $V=0$, combining this controllability
property   with the result of Beauchard \cite{BCH}, we see that
the system is globally exactly controllable in
 $ S\cap H^{5+\e}$
generically with respect to $Q\in C^\ty(\overline{D},\R)$, i.e.,
 for any $z_0,z_1\in S\cap H^{5+\e}$ there is a time $T>0$
and a control $u\in H_0^1([0,T],\R)$ such that $z(0,u)=z_0$ and
$z(T,u)=z_1$.

Next we apply approximate controllability property to prove that
the random Schr\"odinger equation has at most one stationary
measure on $S$. This follows from
  uniform Feller property  and irreducibility of the transition functions of the Markov chain associated to the system in question.
   Existence of a stationary probability different from the Dirac measure concentrated at zero is an open problem.
   There are several results on the existence of stationary measures for   deterministic   Schr\"odinger equations.
Bourgain \cite{Bou1} and Tzvetkov \cite{NTZ1} prove the existence
of   stationary Gibbs measures for different   nonlinear
Schr\"odinger systems. Kuksin and Shirikyan \cite{KUKSHIR}
construct a stationary measure as a limit  of the unique
stationary measure  of the randomly forced complex
Ginzburg--Landau equation when the viscosity goes to zero. In
\cite{DOd}, Debussche and Odasso prove existence and uniqueness of
stationary measure and a polynomial mixing property for a damped
1D Schr\"odinger equation. For finite-dimensional approximations
of the Schr\"odinger equation, existence and uniqueness of
stationary measure and an exponential mixing
 property    is obtained in~\cite{N}.

\textbf{Acknowledgments.} The author would like to thank Armen
Shirikyan for many valuable discussions and support. In
particular, for the remark that the controllability property
implies   uniqueness of stationary measure for the random
Schr\"odinger equation.

\vspace{52pt} \textbf{Notation}\\\\ In this paper we use the
following notation. Let $D \subset \R^m, m\ge1$ be a bounded
domain with smooth boundary. Let  $H^s:=H^s(D)$  be the Sobolev
space of order $s\ge0$ endowed with the norm $\|\cdot\|_s$.
Consider the operators $ -\Delta z +Vz,z\in\DD(- \Delta +V):=H_0^1
\cap H^2, $ where $V\in C^\ty(\overline{D},\R)$. We denote by
$\{\la_{j,V} \}$ and $\{e_{j,V} \}$ the sets of eigenvalues and
normalized eigenfunctions  of $ -\Delta+V$. Define the spaces
$H^s_{(V)}:=D((-\Delta+V)^\frac{s}{2})$. Let
$\langle\cdot,\cdot\rangle$ and $\|\cdot\|$ be the scalar product
and the norm in the space $L^2 $. Let $S$ be the unit sphere in
$L^2 $. For a Polish space $X$, we shall use the following
notation.
\\ $\BBBBB(X)$ is the $\sigma$-algebra of Borel subsets of $X$.
\\$C(X)$ is the space of real-valued continuous functions on $X$.
\\ $C_b(X)$ is the space of bounded functions $f\in C(X)$.
\\$\LL(X)$ is the space of functions $f\in C_b(X)$ such that
$$
\|f\|_{\LL}:=\|f\|_\infty+\sup_{u\neq
v}\frac{|f(u)-f(v)|}{\|u-v\|}<+\infty.
$$$\PP(X)$ is the set of probability measures on $(X,\BBBBB(X)).$
\section{Main results}\label{S:MR}
\subsection{Approximate controllability to $e_{1,V}$}\label{S:APROX}
   The following lemma shows the well-posedness of
system (\ref{E:hav1}), (\ref{E:ep1}).

\begin{lemma}\label{L:lav}
 For any   control   $ u\in C^\ty([0,\ty),\R)
 $ such that $\frac{\dd^k  u}{\dd t^k}(0)=0$ for all
  $k\ge1 $ and for any    $ z_0\in H^s_{(V+ u(0)Q) } $
  problem
(\ref{E:hav1})-(\ref{E:sp1}) has a unique solution  $ z\in C([0,\infty),H^s) $. 
Furthermore, if  $  \frac{\dd^k  u}{\dd t^k}(T)=0$ for all $k\ge1
$,  then $ z(T)\in ~H^s_{(V+ u(T)Q)}$.
\end{lemma}
See  \cite{BCH} for the proof. Denote by $\UU_t(\cdot,u):
L^2\rightarrow L^2 $ the resolving operator of (\ref{E:hav1}),
(\ref{E:ep1}).
 As in \cite{VN}, we assume that the functions $V$
and $Q$ satisfy the following condition.

\begin{condition}\label{C:p}
The functions $V,Q\in C^\ty(\overline{D},\R) $ are such that:
\begin{enumerate}
\item [(i)] $\langle Qe_{1,V},e_{j,V}\rangle\neq0$
  for all $j\ge2$,
\item  [(ii)]$\la_{1,V}-\la_{j,V}\neq \la_{p,V}-\la_{q,V}$ for all  $j,p,q\ge1$ such that
  $\{1,j\}\neq\{p,q\}$ and $j\neq 1$.
\end{enumerate}
\end{condition}
 We say that problem (\ref{E:hav1}), (\ref{E:ep1}) is
approximately controllable to $e_{1,V}$ in $H^s$ if for any
integer $k\ge1$, any constants $\e,\de,d>0$ and for any point $z_0
\in S\cap H^s_{(V)}$ there is a time $T>0$ and a control $u\in
C_0^\ty((0,T),\R)$, $\|u\|_{C^k([0,T])}<d$ such that
\begin{equation}
\|\UU_T(z_0,u)-e_{1,V}\|_{s-\de}<\e.\nonumber
\end{equation}
The theorem below  is one of the main results of the present
paper.
 \begin{theorem}\label{T:stab}
 Under Condition \ref{C:p}, for any $s>0$, problem (\ref{E:hav1}), (\ref{E:ep1}) is
approximately controllable to $e_{1,V}$ in $H^s$.
 \end{theorem}
In many relevant examples, the spectrum of the operator
$-\Delta+V$ is degenerate, hence property (ii) in Condition
\ref{C:p} is not verified. In fact, the proof of Theorem
\ref{T:stab} can be adapted to show  the approximate
controllability of   (\ref{E:hav1}), (\ref{E:ep1})
  for any potential $V$, under stronger assumptions on the function $Q$.
More precisely, we have the following result.
\begin{theorem}\label{T:aaa}
Let  $V\in C^\ty(\overline{D},\R)$ be an arbitrary function. The
set of potentials $Q$ such that problem (\ref{E:hav1}),
(\ref{E:ep1}) is approximately controllable to $e_{1,V}$ in $H^s$
for any $s>0$ is residual in $C^\ty(\overline{D},\R)$, i.e.,
contains a countable intersection of open dense sets in
$C^\ty(\overline{D},\R)$.
 \end{theorem}See Subsection~\ref{S:STAB} for the proof this theorem.

 We say that problem (\ref{E:hav1}), (\ref{E:ep1}) is
approximately controllable in $L^2$  if for any integer $k\ge1$,
any constants $\e,d>0$ and for any points $z_0,z_1\in S$ there is
a time $T> 0$ and a control  $u\in C_0^\ty((0,T),\R)$,
$\|u\|_{C^k([0,T])}<d$  such that
\begin{equation}
\|\UU_k(z_0,u)-z_1\|<\e.\nonumber
\end{equation}Combination of Theorem \ref{T:aaa} with the time reversibility property of the Schr\"odinger equation implies approximate controllability in $L^2$.
\begin{theorem}\label{T:contr}Let  $V\in C^\ty(\overline{D},\R)$ be an arbitrary function. The set of potentials $Q$ such that problem (\ref{E:hav1}),
(\ref{E:ep1}) is approximately controllable in $L^2$
 is residual in $C^\ty(\overline{D},\R)$.
\end{theorem}This result is proved exactly in the same way as Theorem 3.5 in \cite{VN}.
 \begin{proof}[Proof of Theorem \ref{T:stab}] By Lemma 3.4 in \cite{VN}, it suffices to prove the theorem for any  initial data $z_0\in S\cap H^s_{(V)} $   with
$\lag z_0,e_{1,V} \rag\neq0$. Let us  introduce the Lyapunov
function
\begin{equation}\label{E:ARE}
\VV(z): =\alpha \|( -\Delta+ V  )^{\frac{s}{2}}P_{1,V}z\|
^2+1-|\lag z,e_{1,V} \rag |^2,\,\,\,\, z\in S\cap H^s_{(V)},
\end{equation}where
 $\alpha>0$
  and   $P_{1,V}z:=z-\lag z,e_{1,V}\rag e_{1,V}$ is  the
orthogonal projection in $L^2$ onto the closure of the vector span
of $\{e_{k,V}\}_{k\ge 2}$. Notice that $\VV(z)\ge0$ for all $ z\in
S\cap H^s_{(V)} $ and $\VV(z)=0$ if and only if $z=ce_{1,V},
|c|=1$. For any $ z\in S\cap H^s_{(V)} $  we have
\begin{align*}
\VV(z)\ge\alpha \|( -\Delta+{V} )^{\frac{s}{2}}P_{1,V}z\| ^2\ge
C_1\|       z \| ^2_s-C_2 .
\end{align*}Thus
\begin{equation}\label{E:sah1}
C(1+\VV(z))\ge \|z\|_s
\end{equation}for some constant $C>0$.
We need the following result  proved in Subsection~\ref{S:HIMTA}.

\begin{proposition}\label{L:hav} There is a finite  or countable set $J\subset \R_{+}^*$ such that for any
  $\alpha\notin J$ and for any $z_0\in S\cap H^s_{(V)} $   with
$\lag z_0,e_{1,V} \rag\neq0$
  and $0<\VV(z_0) $
  there is a  time $T>0$ and a control~$u\in
C_0^\ty((0,T),\R)$, $\|u\|_{C^k([0,T])}<d$ verifying
\begin{equation*}
\VV(\UU_T(z_0,u))<\VV(z_0).
\end{equation*}\end{proposition}

Let us take any  $z_0\in S\cap H^s_{(V)} $   with $\lag
z_0,e_{1,V} \rag\neq0$
  and $0<\VV(z_0) $, and choose $\alpha>0$ in (\ref{E:ARE}) such that $\VV(z_0)<1$.
Define the set
\begin{align*}
\KK=\{z\in   H^{s}_{(V)} : \UU_{T_n} (z_0,u_{n}) \rightarrow z
\,\,\,&\text{in $H^{s-\de}$}
  \,\,\text{for some}\,\,  T_n\ge0
\,\,\, \\&  \text{and}\,\, u_n \in C^\ty_0( (0,T_n),\R),
\|u_n\|_{C^k([0,T_n])}<d \}.\nonumber
\end{align*}
Let
$$
m:=\inf_{z\in\KK}\VV(z).
$$
This infimum is attained, i.e., there is $e\in \KK$ such that
\begin{equation}\label{E:labz1}
\VV(e)=\inf_{z\in\KK}\VV(z).
\end{equation}Indeed, take any minimizing sequence $z_n\in\KK$, $\VV(z_n)\rightarrow
m$. By (\ref{E:sah1}), $z_n$ is bounded in $H^s$. Thus, without
loss of generality, we can assume that   $z_n\rightharpoonup e$
in~$H^{s}$ for some $e\in H^s_{(V)}$. This implies that
$\VV(e)\le\liminf_{n\rightarrow\ty}\VV(z_n)= m$. Let us show that
$e\in \KK$. As $z_n\in\KK$, there are sequences $T_n>0$ and
$u_n\in C_0^\ty((0,T_n),\R)$ such that
\begin{align}
\|\UU_{T_n} (z_0,u_n )-z_n\|_{s-\de}
&\le\frac{1}{n}.\label{E:ppp1}
\end{align}
On the other hand,   $z_n\rightarrow e$ in $H^{s-\de}$, and
(\ref{E:ppp1}) implies that $\UU_{T_n} (z_0,u_{n})) \rightarrow e$
in $H^{s-\de}$. Thus $e\in\KK$ and  $\VV(e)= m$.

Let us show that $\VV(e)=0.$ Suppose, by contradiction, that
$\VV(e)>0$. It follows from (\ref{E:labz1}) and from the choice of
$\alpha$ that
 $\VV(e)\le\VV(z_0)<1$. This shows that $\lag e,e_{1,V} \rag\neq0$.
 Proposition \ref{L:hav} implies that
       there is a time $T>0$ and a control
$u\in C^\ty_0((0,T),\R)$  such that
\begin{equation}\label{E:ap11}
\VV(\UU_{T} (e,u))<\VV(e).
\end{equation}
Define $\tilde{u}_n(t)=u_n(t)$, $t\in[0,T_n]$ and
$\tilde{u}_n(t)=u (t-T_n)$, $t \in[T_n,T_n+T]$.  Then
$\tilde{u}_n\in C^\ty_0((0,T_n+T),\R)$ and, by continuity in
$H^{s-\de}$ of the resolving operator for (\ref{E:hav1}),
(\ref{E:ep1}) (e.g., see \cite{CW}),
$$\UU_{T_n+T}  (z_0,\tilde{u}_n)\rightarrow
\UU_{T} (e,u)\,\,\,\text{in $H^{s-\de}.$}
$$ This implies that
$\UU_{T} (e,u  )\in \KK$. Clearly, (\ref{E:ap11}) contradicts
(\ref{E:labz1}). It follows that $\VV(e)=0$, hence $ e=ce_{1,V} $
for some $c:=c(e)\in\C$, $|c|=1$. As
$\UU_{\tau}(ce_{1,V},0)=e^{i\tau}ce_{1,V}=e_{1,V}$ for
$\tau=-\arg(c)$, we see that $e_{1,V}\in\KK$.

\end{proof}

\subsection{Proof of Proposition \ref{L:hav}}\label{S:HIMTA}
 Take any $z_0\in S\cap H^s_{(V)} $   such that
$\lag z_0,e_{1,V} \rag\neq0$
  and $0<\VV(z_0) $. This implies that also $\lag z_0,e_{p,V} \rag\neq0$ for some $p\ge2$. Let us
consider the mapping
\begin{align*}
\VV(\UU_T (z_0,(\cdot)w)) :  \R &\rightarrow\R, \\
\sigma&\rightarrow\VV(\UU_T (z_0,\sigma w)),
\end{align*}where $T>0$ and $w\in C^\ty_0((0,T),\R)$.
We are going to show that,   for an appropriate choice of $T$ and
$w$, we have $\frac{\dd \VV(\UU_T (z_0,\sigma w))}{\dd
\sigma}|_{\sigma=0}\neq 0$. Clearly,
\begin{align}\label{E:ccc11}
\frac{\dd\VV(\UU_T (z_0,\sigma w))}{\dd
\sigma}\Big|_{\sigma=0}&=2\alpha\Re\lag (-\Delta+ V
)^\frac{s}{2}P_{1,V}\UU_T (z_0,0),\nonumber\\&\quad\quad(-\Delta+
V )^\frac{s}{2}P_{1,V}\RR_T(w)\rag\nonumber\\&\quad-2\Re\lag\UU_T
(z_0,0),e_{1,V}\rag\lag e_{1,V},\RR_T(w)\rag,
\end{align}where $\RR_t(\cdot)$ is the resolving operator of problem
\begin{align}
i {\dot z}  &= -\Delta z+ V(x)z+ w(t) Q(x)\UU_t(z_0,0),\,\,\,\,x\in D,\label{E:hav191}\\
z\arrowvert_{\partial D}&=0, \label{E:ep191}\\
 z(0 )&=0.\label{E:sp191}
\end{align}System (\ref{E:hav191})-(\ref{E:sp191})  is the linearization of (\ref{E:hav1}),
(\ref{E:ep1}) around the solution $\UU_t (z_0,0)$. Note that
(\ref{E:hav191})-(\ref{E:sp191}) is equivalent to
\begin{equation}\label{E:1D1}
z(t)= -i\int_0^t e^{-i(-\Delta+ V)(t-\tau)}w(\tau) Q(x)\UU_\tau
(z_0,0)\dd \tau.
\end{equation}Taking into account the fact that
\begin{equation}\label{E:aaa11}
\UU_t (z_0,0)=\sum_{k=1}^\ty e^{-i\la_{k,V}t}\lag z_0, e_{k,V}\rag
e_{k,V},
\end{equation}we get from (\ref{E:1D1})
\begin{equation}\label{E:bbb11}
\lag \RR_T(w), e_{j,V}\rag=-ie^{-i\la_{j,V}T}\sum_{k=1}^\ty\lag
z_0, e_{k,V}\rag \lag Qe_{k,V},e_{j,V}\rag\int_0^T
 e^{-i(\la_{k,V}-\la_{j,V})\tau} w(\tau)  \dd \tau.
\end{equation}Replacing (\ref{E:aaa11}) and (\ref{E:bbb11}) into
(\ref{E:ccc11}), we get
\begin{align*}
\frac{\dd\VV(\UU_T (z_0,\sigma w))}{\dd
\sigma}\Big|_{\sigma=0}&=-2\alpha\Im\sum_{j=2,k=1}^\ty \la_{j,V}^s
\lag z_0, e_{j,V}\rag \lag e_{k,V},z_0 \rag \lag
Qe_{k,V},e_{j,V}\rag\nonumber\\&\quad\quad\times\int_0^T
 e^{ i(\la_{k,V}-\la_{j,V})\tau} w(\tau)  \dd \tau\nonumber\\&\quad+2
 \Im\sum_{ k=1}^\ty    \lag z_0,
e_{1,V}\rag \lag e_{k,V},z_0 \rag \lag
Qe_{k,V},e_{1,V}\rag\nonumber\\&\quad\quad\times\int_0^T
 e^{ i(\la_{k,V}-\la_{1,V})\tau} w(\tau)  \dd \tau\\&=\int_0^T\Phi(\tau)w(\tau)\dd\tau,
\end{align*}where
\begin{align}\label{E:eee1}
i\Phi(\tau):&=-\alpha \sum_{j=2,k=1}^\ty   \la_{j,V}^s \lag z_0,
e_{j,V}\rag \lag e_{k,V},z_0 \rag \lag Qe_{k,V},e_{j,V}\rag
 e^{ i(\la_{k,V}-\la_{j,V})\tau}  \nonumber\\&\quad+\alpha \sum_{j=2,k=1}^\ty   \la_{j,V}^s \lag e_{j,V},z_0
\rag \lag z_0,e_{k,V} \rag \lag Qe_{k,V},e_{j,V}\rag
 e^{- i(\la_{k,V}-\la_{j,V})\tau}   \nonumber\\&\quad+
  \sum_{ k=2}^\ty    \lag z_0,
e_{1,V}\rag \lag e_{k,V},z_0 \rag \lag Qe_{k,V},e_{1,V}\rag
 e^{ i(\la_{k,V}-\la_{1,V})\tau}     \nonumber\\&\quad-
  \sum_{ k=2}^\ty    \lag e_{1,V},z_0
\rag \lag z_0,e_{k,V} \rag \lag Qe_{k,V},e_{1,V}\rag
  e^{ -i(\la_{k,V}-\la_{1,V})\tau}\nonumber\\&=\sum_{j=2,k=2}^\ty
P(z_0,Q,j,k) e^{-i(\la_{j,V} -\la_{k,V} )t}
\nonumber\\&\quad+\sum_{j=2}^\ty  (\alpha\la_{j,V}^s+1)\langle
z_0,e_{1,V} \rangle \langle e_{j,V} ,z_0\rangle   \langle Qe_{j,V}
,e_{1,V}\rangle  e^{-i(\la_{1,V} -\la_{j,V} )t}
\nonumber\\&\quad-\sum_{j=2}^\ty (\alpha\la_{j,V}^s+1)\langle
e_{1,V}, z_0\rangle \langle  z_0,e_{j,V}\rangle   \langle Qe_{j,V}
,e_{1,V}\rangle  e^{i(\la_{1,V} -\la_{j,V} )t}.
\end{align} Here $P(z_0,Q,j,k)$ are constants. Define the set
\begin{align*}J:=\{\alpha\in\R:\alpha\la_{j,V}^s=-1\,\,\,  \text{for some
$j\ge1$}\}, \end{align*} and take any   $\alpha\notin J$.  As
$\lag z_0,e_{j,V}\rag\neq0$ for $j=1,p$, Condition \ref{C:p} and
Lemma~3.10 in \cite{VN} imply that there is $T>0$ such that
$\Phi(T)\neq 0$. Thus there is  a control $w\in
C^\ty_0((0,T),\R)$, $\|w\|_{C^k([0,T])}<d$ such that $$
\frac{\dd\VV(\UU_T (z_0,\sigma w))}{\dd
\sigma}\Big|_{\sigma=0}=\int_0^T\Phi(\tau)w(\tau)\dd\tau\neq0.$$
This implies that there is $\sigma_0\in\R$ close to zero such that
$$
\VV(\UU_T (z_0,\sigma_0 w))<\VV(\UU_T (z_0,0))=\VV( z_0 ),
$$which completes the proof.

\begin{remark} Modifying slightly
   Condition \ref{C:p}, Theorem \ref{T:stab} can be restated for    the  eigenfunction $e_{i,V}$,  $i\ge1$. Indeed,
   one should replace $\la_{1,V}$ and $e_{1,V}$ by $\la_{i,V}$  and
   $e_{i,V}$ in Condition \ref{C:p} and use the   Lyapunov function
\begin{equation*}
\VV_i(z): =\alpha \|( -\Delta+ V  )^{\frac{s}{2}}P_{i,V}z\|
^2+1-|\lag z,e_{i,V} \rag |^2,\,\,\,\, z\in S\cap H^s_{(V)},
\end{equation*}where
    $P_{i,V}$ is  the
orthogonal projection in $L^2$ onto the closure of the vector span
of $\{e_{k,V}\}_{k\neq i}$.

\end{remark}

\subsection{Proof of Theorem \ref{T:aaa}}\label{S:STAB}
Let us look for a control in the form $\sigma+u(t)$, where
$\sigma\in\R\backslash\{0\}$ is a constant. Then (\ref{E:hav1})
takes the form
$$i\dot z    = -\Delta z+(V(x)+ \sigma Q(x))z+ u(t) Q(x)z,
$$and the idea of the proof is to show that the set of potentials $Q$ such that Condition \ref{C:p} holds for $V,Q$ replaced by $V + \sigma Q, Q$ is residual.
The  proof is divided into   three steps.

 \vspace{6pt}\textbf{Step 1.} First let us show that the set $\QQ$ of all functions  $Q$ verifying
\begin{equation}\label{E:epae}
\la_{1,Q}-\la_{j,Q}\neq \la_{p,Q}-\la_{q,Q}
\end{equation}
for all integers $j,p,q\ge1$   such that
  $\{1,j\}\neq\{p,q\}$ and $j\neq 1$ is residual in $
C^\ty(\overline{D},\R)$.
  Fix $j,p,q\ge1$ and denote by $\QQ_{j,p,q}$ the set of   functions $Q\in
C^\ty(\overline{D},\R)$ satisfying (\ref{E:epae}). As
$\QQ=\cap_{j,p,q} \QQ_{j,p,q}$, it suffices to show that
$\QQ_{j,p,q}$ is open and dense. Continuity of the eigenvalues
$\la_{k,Q}$     from $ C^\ty(\overline{D},\R)$ to $\R$ (e.g.,
see~\cite{KATO}) implies that  $\QQ_{j,p,q}$ is open. Let us show
that
 $\QQ_{j,p,q}$ is dense in $ C^\ty(\overline{D},\R)$. Indeed, by \cite{ALBERT}, the set $\widetilde{\QQ }$ of functions $Q\in
C^\ty(\overline{D},\R)$
   such that the spectrum of the operator $-\Delta+Q$ is  non-degenerate  is residual in $ C^\ty(\overline{D},\R)$. Take any $Q\in \widetilde{\QQ } $ and $P\in C^\ty(\overline{D},\R)$.
    It is well known that $\la_{k,Q+\sigma  P}$ and  $e_{k,Q+\sigma  P}$ are  analytic in $\sigma$ in the neighborhood of $0$ in $\R$ (e.g., see \cite{KATO}).
 Differentiating the identity
$$(-\Delta+Q +\sigma P-\la_{k,  Q+\sigma P})e_{k,  Q+ \sigma P}=0$$
with respect to $\sigma$ at $\sigma=0$, we get
$$
(-\Delta+Q -\la_{k, Q})\frac{\dd e_{k,Q+\sigma
 P} }{\dd
\sigma}\Big|_{\sigma=0}+(P-\frac{\dd \la_{k,Q+\sigma P } }{\dd
\sigma}\Big|_{\sigma=0})e_{k,Q }=0.
$$Taking the scalar product of this identity with $e_{k,Q}$, we
obtain
\begin{equation}\label{E:ppa}
\frac{\dd \la_{k,Q+\sigma
 P} }{\dd
\sigma}\Big|_{\sigma=0}= \langle P,  e_{k,Q} ^2  \rangle.
\end{equation}
Thus
\begin{equation}\label{E:ppap}
\frac{\dd  }{\dd \sigma}(\la_{1,Q+\sigma
 P}-\la_{j,Q+\sigma
 P}-\la_{p,Q+\sigma
 P}+\la_{q,Q+\sigma
 P})\Big|_{\sigma=0}= \langle P,  e_{1,Q} ^2- e_{j,Q} ^2- e_{p,Q} ^2+ e_{q,Q} ^2  \rangle.
\end{equation}
By Theorem \ref{T:thm}, the  set $\aA$ of all functions $Q$ such
that the system $\{e_{j,Q}^2\}_{j=1}^\ty$ is rationally
independent is residual    in $C^\ty(\overline{D},\R)$. Hence the
set $\widetilde{\QQ} \cap \aA$ is residual. Fix any $Q\in
\widetilde{\QQ} \cap \aA$
 and take $P\in C^\ty(\overline{D},\R)$ such that
$$
\langle P,  e_{1,Q} ^2- e_{j,Q} ^2- e_{p,Q} ^2+ e_{q,Q} ^2
\rangle\neq0.
$$Then (\ref{E:ppap}) implies that $Q+\sigma
 P\in\QQ_{j,p,q}$ for any $\sigma$ sufficiently close to $0$. This proves that $\QQ_{j,p,q}$ is dense in $C^\ty(\overline{D},\R)$.

 \vspace{6pt}\textbf{Step 2.} Take any sequence $\sigma_n \rightarrow 0$, $\sigma_n\neq 0$. Then the set $\PP_1$ of all functions
 $Q\in C^\ty(\overline{D},\R)$
such that $V+\sigma_n Q\in \QQ$ for all $n\ge1$ is residual as a
countable intersection of residual sets. On the other hand, the
set $\PP_2$ of   functions $Q\in C^\ty(\overline{D},\R)$ such that
$\langle Qe_{1,V},e_{j,V}\rangle\neq0$ for all $j\ge2$ is also
residual (see Section 3.4 in \cite{VN}). Define
$\PP:=\PP_1\cap\PP_2$.
  Let us show that problem
(\ref{E:hav1}), (\ref{E:ep1}) is approximately controllable to
$e_{1,V}$ for any $Q\in\PP$. Fix an integer $n\ge1$ and consider
the problem
\begin{align}
i\dot z   &= -\Delta z+V(x)z+\sigma_nQ(x)z+ u(t) Q(x)z,\,\,\,\,x\in D,\label{E:hav92}\\
z\arrowvert_{\partial D}&=0,\label{E:ep92}\\
 z(0,x)&=z_0(x).\label{E:sp92}
\end{align} Let $\UU_t^n(\cdot,u)$ be the resolving operator
for problem    (\ref{E:hav92}), (\ref{E:ep92}). Then we have
$\UU_t^n(\cdot,u)=\UU_t  (\cdot,u+\sigma_n)$.  Notice that we
cannot apply Theorem \ref{T:stab} with potentials
 $V(x)+\sigma_nQ(x)$ and $Q$. Indeed, Condition \ref{C:p}, (i)
is not necessarily satisfied. However, as $\sigma_n\rightarrow0$
and $Q\in\PP_2$, for any $N\ge1$ there is $n^*\ge1$ such that
$\langle Qe_{1,V+\sigma_nQ},e_{j,V+\sigma_n Q}\rangle\neq0$ for
any $j=1,\ldots,N$ and $n\ge n^*$.
 Modifying slightly the arguments of the proof of Theorem \ref{T:stab},
   we show that  this property is enough to conclude the approximate controllability.

We need the following result.
\begin{lemma}\label{L:po}
For any  $M>0$ and $\e>0$ there is an integer
$\hat{n}=\hat{n}(M,\e,Q)\ge~1$ such that for any  $n\ge \hat{n} $
  and $z_0\in S\cap H^{s}_{(V+\sigma_n Q)}$
with $\langle z_0,e_{1,V+\sigma_n Q}\rangle\neq0$ and $\|z_0\|_{s
}<M$ we have
$$ \|\UU_{T }^n(z_0,u)-e_{1,V+\sigma_n Q}\|_{s-\de}<\e
$$ for some time $T>0$ and control $u\in C^\ty_0((0,T),\R),\|u\|_{C^k([0,T])}<d$.
\end{lemma} To prove Theorem
\ref{T:aaa}, let $z_0\in S\cap H^{s}_{(V)}$ be such that
$\|z_0\|_{s }<M$ and $\langle z_0,e_{1,V  }\rangle\neq0$. As $z_0$
is not necessarily in $H^{s}_{(V+\sigma_n Q)}$, we cannot apply
 Lemma~\ref{L:po}.
Take any $v\in C^\ty ([0,\eta],\R)$ such that $\frac{\dd^k v}{\dd
t^k}(0)=\frac{\dd^k v}{\dd t^k}(\eta)=0$ for all
  $k\ge 1  $ and $v(0)=0$.
By Lemma~\ref{L:lav}, we have   $y:=\UU_{\eta} (z_0,v)\in
H^s_{(V+v(\eta)Q)}$. If $k\ge1$ is sufficiently large and
$\|v\|_{C^k
  ([0,\eta] )}$ is sufficiently small, then $\|y\|_{s
  }<M$   and $\lag y, e_{1,V+v(\eta)Q}\rag\neq 0$. We can choose $v$ such that $v(\eta)=\sigma_n$ for some $n\ge\hat{n}$.
   Applying Lemma~\ref{L:po} for the initial data $y$, we   see
  that there is a time $  \tilde{T}>0$ and a  control $\tilde u $ such that
   $\tilde u-v(\eta)\in C^\ty_0 ((0,\tilde
  T),\R))$  and
  $$ \|\UU_{\tilde T }  (y,\tilde u)-e_{1,  V+ v{(  \eta)Q}}\|_{s-\de }<\frac{\e}{2}.
$$ For sufficiently small  $  \|v\|_{C^k
  ([0,\eta] )}  +\eta$ we have
   $$ \|\UU_{\eta  }  (\UU_{\tilde T }   (y,\tilde u),v(\eta-\cdot))-e_{1,V
   }\|_{s-\de
   }<\e.
$$
  This proves Theorem \ref{T:aaa}.

\vspace{6pt}\textbf{Step 3.} To prove Lemma \ref{L:po}, let $\VV$
be defined by (\ref{E:ARE}) with $V+\sigma_n Q$ instead of $V$. As
$\langle z_0,e_{1,V+\sigma_n Q}\rangle\neq0$, we can choose
$\alpha>0$ so small that $\VV(z_0)<1$. Notice that if $ \VV(z)<1$,
$z\in S$ then $\lag z,e_{1,V+\sigma_n Q}\rag\neq0.$
 On the other hand, for any $\e>0$  there is an integer $N\ge1$
such that if $ \|z\|_s\le M$, $z\in S\cap H^{s}_{(V+\sigma_n Q)}$
and $\lag z,e_{j,V+\sigma_n Q}\rag=0$ for any $j\in[2,N]$, then
$\|z-ce_{1,V+\sigma_n Q}\|_{s-\de}<\e$ for some $c:=c(z)\in\C$
with $|c|=1$.

We need the following lemma.

\begin{lemma}
\label{L:ppp}  There is a finite  or countable set $J\subset
\R_{+}^*$ and  an integer
  $\hat{n} \ge1$ such that for any
  $\alpha\notin J$,  $n\ge \hat{n}  $  and     $z\in H^{s}_{(V+\sigma_n Q)}$ with  $\lag
z,e_{1,V+\sigma_n Q}\rag\neq0$  and  $\lag z,e_{j,V+\sigma_n
Q}\rag\neq0$ for some integer $j\in[2,N]$, there is a time $T>0$
and a control $u\in C^\ty_0((0,T),\R)$   verifying
$$
\VV(\UU_{T}^n(z,u))<\VV(z).
$$
\end{lemma}\begin{proof} As $Q\in\PP$, for sufficiently large  $\hat{n} $ we have $\langle Qe_{1,V+\sigma_nQ},e_{j,V+\sigma_nQ}\rangle\neq0$
for   $j=2,\ldots,N$ and $n\ge\hat{n}$.   Repeating the arguments
of the proof of    Proposition~\ref{L:hav}, one should just notice
that if sum (\ref{E:eee1}) equals   zero for all $\tau\ge 0$, then
$\lag z_0,e_{j,V+\sigma_n Q}\rag=0$ for all $j\in[2,N]$.
 This contradicts the hypothesis of the lemma.\end{proof}
As in the proof of Theorem \ref{T:stab}, we define the set
\begin{align*}
\KK=\{z\in   H^{s}_{(V+\sigma_n Q)} : \UU_{T_\ell}^n(z_0,u_{\ell})
&\rightarrow z \,\,\,\text{in $H^{s-\de}$}
  \,\,\text{for some}\,\,  T_\ell\ge0,
\,\,\, \\&  \text{and}\,\, u_\ell \in C^\ty_0(
(0,T_\ell),\R),\|u_\ell\|_{C^k([0,T_\ell])}<d \}.\nonumber
\end{align*}
Let
$$
m:=\inf_{z\in\KK}\VV(z).
$$
This infimum is attained at some point $e\in \KK$. Let us show
that $$\|e-ce_{1,V+\sigma_n Q}\|_{s-\de}<\e$$ for some
$c:=c(z)\in\C$ with $|c|=1$. Indeed, it follows from
  the choice of $\alpha$ that
 $\VV(e)\le\VV(z_0)<1$. Hence $\lag e,e_{1,V+\sigma_n Q} \rag\neq0$.
 Suppose that $\lag e,e_{j,V+\sigma_n Q}\rag\neq0$ for
some integer $j\in[2,N]$. By Lemma
 \ref{L:ppp},      there is a time $T>0$ and a control
$u\in C^\ty_0((0,T),\R)$  such that
\begin{equation}\label{E:ap1}
\VV(\UU_{T}^n(e,u))<\VV(e).
\end{equation}
Define $\tilde{u}_\ell(t)=u_\ell(t)$, $t\in[0,T_\ell]$ and
$\tilde{u}_\ell(t)=u (t)$, $t \in[T_\ell,T_\ell+T]$.  Then
$\tilde{u}_\ell\in C^\ty_0((0,T_\ell+T),\R)$ and
$$\UU_{T_\ell+T}^n(z_0,\tilde{u}_\ell)\rightarrow
\UU_{T}^n(e,u)\,\,\,\text{in $H^{s-\de}.$}
$$ This implies that
$\UU_{T}^n(e,u)\in \KK$. Clearly, (\ref{E:ap1}) contradicts the
definition of $e$. It follows that $\lag e,e_{j,V+\sigma_n
Q}\rag=0$ for any  $j\in[2,N]$, hence $\|e-ce_{1,V+\sigma_n
Q}\|_{s-\de}<\e$ for some $c:=c(e)\in\C$, $|c|=1$.
  Without loss of generality, we can
suppose  that  $c=1$.

\section{Applications}

\subsection{Global exact controllability}\label{S:ppp}

 The following  controllability result  for system
(\ref{E:hav1}), (\ref{E:ep1}) is obtained by  Beau\-chard  in the
case $m=1$ and $V=0$.
 \begin{theorem}\label{T:BCH1}
There is a    residual set $ \QQ' $  in $C^\ty(\overline{D},\R)$
such that for any $Q\in  { {\QQ'}}$ and some constants  $T>0$ and
$\e>0$ the following exact controllability property holds:  for
any $z_0,z_1\in S\cap H^{5+\de}_{(0)}$ with
$$
\|z_i-e_{1,0}\|_{5+\de}<\e,\,\,\,i=1,2
$$
there is a control $u\in H_0^1([0,T],\R)$ such that
\begin{equation}
\UU_T(z_0,u)=z_1.\nonumber
\end{equation}
\end{theorem}See \cite{BCH} for the proof in the case $Q(x)=x$. The existence of a  residual set $ \QQ' $ is proved in \cite{BPE}, using the same
ideas as in \cite{BCH}.  Combining Theorems \ref{T:aaa} and
\ref{T:BCH1}, we obtain.
 \begin{theorem}\label{T:BV}
There is a    residual set $ \widehat{{ {\QQ}}}$  in
$C^\ty(\overline{D},\R)$ such that for any $Q\in \widehat{{
{\QQ}}}$ and any $z_0,z_1\in S\cap H^{5+\de}_{(0)}$ there is a
time $T>0$  and a control $u\in H_0^1([0,T],\R)$ verifying
\begin{equation}
\UU_T(z_0,u)=z_1.\nonumber
\end{equation}
\end{theorem}

\begin{remark}
It is shown in  \cite{BCH,BM} that if we take $Q(x)=x$, then for
any $N\ge1$ there is a constant  $\sigma^*>0$ such that for any
$\sigma\in(0,\sigma^*)$ we have
\begin{enumerate}
\item [$(i)$] $\langle x e_{1,\sigma x},e_{j,\sigma x}\rangle\neq0$
  for all $j\ge2$,
\item  [$(ii)$]$\la_{1,\sigma x}-\la_{j,\sigma x}\neq \la_{p,\sigma x}-\la_{q,\sigma x}$ for all  $j,p,q\ge1$ such that
  $\{1,j\}\neq\{p,q\}$ and $2\le j\le N$.
\end{enumerate}
Thus the proof of Theorem \ref{T:aaa} works giving the approximate
controllability.
 On the other
hand, by Beauchard \cite{BCH}, for  $Q(x)=x$  the problem is
exactly controllable in a neighborhood of $e_{1,0}$. Thus $x\in
\widehat{{ {\QQ}}}$.
\end{remark}

\subsection{Uniqueness of stationary measure}
Let us consider the Schr\"odinger equation with a potential which
has a random time-dependent amplitude:
\begin{align}
i\dot z &= -  \Delta z+V(x)z +\beta(t) Q(x)z,\,\,\,\,x\in D,\label{E:havvv}\\
z\arrowvert_{\partial D}&=0,\label{E:epvv}\\
 z(0)&=z_0,\label{E:spvv}
\end{align}where   $V,Q\in C^\ty(\overline{D},\R)$ are   given
functions. We  assume that $\beta(t)$ is a random process of the
form
\begin{equation}\label{E:pu}
\beta (t)=\sum_{k=0}^{+\infty}I_{k}(t)\eta_k(t-k),\,\,\,\,\,t\ge
0,
\end{equation}  where $I_{k}(\cdot)$ is the indicator function of the interval
$[k,k+1)$ and $\eta_k$ are independent identically distributed
 random variables in $L^2([0,1],\R)$. Then $\UU_k(\cdot,\beta)$ is a homogeneous
Markov chain with respect to the filtration $\FF_k$ generated by
$\eta_0,\dots,\eta_{k-1}$(e.g., see \cite{OKS}). For any $z\in S$
and $\Gamma\in \BBBBB (S)$, the transition functions corresponding
to the process $\UU_k(\cdot,\beta)$ are defined by
$P_k(z,\Gamma)=\pP\{\UU_k(z,\beta)\in\Gamma\}$ and the Markov
operators by
$$
\PPPP_k f(z)=\int_SP_k(z,\dd
v)f(v),\,\,\,\,\,\,\,\PPPP^*_k\mu(\Gamma)=\int_S
P_k(v,\Gamma)\mu(\dd v),
$$ where $f\in C_b(S)$ and $\mu\in\PP(S).$ Let us recall that a  measure $\mu\in\PP(S)$ is stationary for
(\ref{E:havvv}), (\ref{E:epvv}), (\ref{E:pu}) if
$\PPPP_1^{*}\mu=~\mu$. The question of  existence of a stationary
measure is an open problem. In this section, we derive the
uniqueness  from Theorem \ref{T:stab}. We need the following
condition.
\begin{condition}\label{C:c}
The random variables $\eta_k$  have the form
$$
\eta_k(t)=\sum_{j=1}^\infty b_j\xi_{jk}g_j(t), \,\,\,\, t\in[0,1],
$$where $\{g_j\}$ is an orthonormal basis in $L^2([0,1],\R)$,
$b_j>0$ are constants with
$$
 \sum_{j=1}^\infty b_j^2<\infty,
$$ and $\xi_{jk}$ are independent real-valued random variables such that
$\E\xi_{jk}^2=1$. Moreover, the distribution of $\xi_{jk}$
possesses a continuous density $\rho_j$ with respect to the
Lebesgue measure and $\rho_j(r)>0$ for all $r\in\R$.
\end{condition}
\begin{theorem}\label{T:MI}
Under Conditions  \ref{C:p} and \ref{C:c},  problem
(\ref{E:havvv}), (\ref{E:epvv}), (\ref{E:pu})  has at most one
stationary measure on $S$.
\end{theorem}
This theorem is derived from the following general result (cf.
\cite{KS2001}). Let $X$ be a Polish space and let $P_k(z,\Gamma)$
be a Markov transition function satisfying the Feller property. We
denote by $\PPPP_k$ and $\PPPP_k^*$
  the corresponding
Markov semigroups. Recall that a stationary measure $\mu$ for
$\PPPP_k^*$
   is said to be ergodic if
   \begin{equation}
\sigma_n(f):=\frac{1}{n}\sum_{i=0}^{n-1}\PPPP_if(z)\rightarrow
(f,\mu)
   \end{equation}for any $f\in C_b(X)$ and for $\mu$-a.e. $z\in X$, where $(f,\mu)=\int_Xf(z)\mu(\dd z) .$
   \begin{theorem}\label{T:Ab}
   Suppose that $P_k$ satisfies the following two conditions.
    \begin{enumerate}
\item[(i)]For any   $f \in \LL(X)$ there is a
constant $L_f > 0$ such that $\PPPP_kf$ is $L_f$-Lipschitz for any
$k\ge 0$.
\item[(ii)] For any point $z \in X$ and any ball $B \subset X$ there is $l \ge 1$ such that
$P_l(z,B) > 0$.
\end{enumerate}Then $\PPPP_k^*$
 has at most one stationary distribution.
   \end{theorem}
\begin{proof} [Proof of Theorem \ref{T:MI}] Let us show that   properties (i) and (ii) are verified for system (\ref{E:havvv}), (\ref{E:epvv}), (\ref{E:pu}).
Take any  function $f \in\LL( S)$. Then
\begin{align*}
 |\PPPP_kf(z_1)- \PPPP_kf(z_2)|&=|\E(f(\UU_k(z_1,\beta))-f(\UU_k(z_2,\beta)))|\\\nonumber&\le\|f\|_{\LL}\E\|\UU_k(z_1,\beta) - \UU_k(z_2,\beta)\|=\|f\|_{\LL}\|z_1-z_2\|,
\end{align*}which implies (i). To show (ii), notice that    Condition \ref{C:c}   implies that
$$
\pP\{\|u-\beta\|_{L^2([0,l])}<\e\}>0
$$for any $u\in L^2([0,l])$
and $\e>0$. Moreover, using the continuity of the mapping
$\UU_l(z_0,\cdot):L^2([0,l])\rightarrow L^2(D)$, for any $\de>0$
we can find   a constant $\e>0$ such that
$$
\pP\{\|\UU_l(z_0,\beta)-\UU_l(z_0,u)\|<\delta\}\ge\pP\{\|u-\beta\|_{L^2([0,l])}<\e\}>0.
$$ Hence, any
point $\UU_l(z_0,u), u\in L^2([0,l])$ is in the support of the
measure $\DD(\UU_l(z_0,\beta))=P_l(z_0,\cdot)$. By Theorem
\ref{T:stab}, problem (\ref{E:hav1}), (\ref{E:ep1}) is
approximately controllable in $S$ (cf. Theorem 3.5 in \cite{VN}),
hence the set $\{\UU_l(z_0,u): u\in L^2([0,l]),l\ge0\}$ is dense
in $S$. This implies (ii). Applying Theorem \ref{T:Ab}, we
complete the proof.
\end{proof}\begin{proof} [Proof of Theorem \ref{T:Ab}]
In view of ergodic decomposition of stationary distributions
(e.g., see \cite{YK}), it suffices to prove that there is at most
one ergodic stationary measure. Let $\mu_1$ and $\mu_2$ be two
ergodic stationary measures. Suppose there is a   function $f
\in\LL( X)$ such that $(f, \mu_1) \neq (f, \mu_2)$. Let $X_i, i =
1, 2$ be the set of convergence in (i) with $\mu = \mu_i$. Then
$\mu_i(X_i) = 1$ and $X_1 \cap X_2 = \emptyset$. Furthermore, in
view of condition (ii), for any ball $B \subset X$ there is $l \ge
1$ such that
$$
\mu_i(B)=\int_X P_l(z,B)\mu_i(\dd z)>0.
$$
Thus $\supp \mu_i = X$, and therefore $\overline{X_i }= X$. Now
let $K_n\subset X$  be an increasing sequence of compact subsets
such that $\mu_i(K_n) > 1 - 2^{-n}$. Then, by condition (i) and
the Arzel\`a theorem, there is a subsequence $k_j \rightarrow \ty$
such that for any $n \ge 1$ the sequence ${\sigma_{k_j} (f)}$
converges uniformly on $K_n$ to an $L_f$-Lipschitz bounded
function $\overline{f}_n$. Let us set $Y = \cup_nK_n$ and define
an $L_f$-Lipschitz function $\overline{f} : Y \rightarrow \R$ such
that $\overline{f}|_{K_n} = \overline{f}_n$. Since $\mu_i(Y ) =
1$, we see that $\mu_i(Y \cap X_i) = 1$ and $\overline{Y \cap X_i}
= X$, where we used again condition (ii). We conclude that
$\overline{f}$ must be a constant function on $X$ equal to $(f,
\mu_i)$. The contradiction obtained shows that $\mu_1 = \mu_2$.

\end{proof}

\section{Independence of squares of eigenfunctions}
Recall that the functions  $\{f_{j}\}_{j=1}^\ty \subset C(D)$ are
said to be rationally independent, if for any $N\ge1$ and
 $\alpha_k\in\Q, k=1,\ldots,N$ with $|\alpha_1|+\cdots+|\alpha_N|>0$ we have
$$\sum_{k=1}^N\alpha_k f_{k}\not\equiv 0.
$$
\begin{theorem}\label{T:thm}
The  set $\aA$ of all functions $Q\in C^\ty(\overline{D},\R)$ such
that the system $\{e_{j,Q}^2\}_{j=1}^\ty$ is rationally
independent is residual    in $C^\ty(\overline{D},\R)$.
\end{theorem}
\begin{proof}The proof of this theorem is inspired by the paper
\cite{PISI}  by Privat and Sigalotti, where the linear
independence of the squares of the eigenfunctions of the Dirichlet
Laplacian     is established to hold generically with respect to
the domain~$D$.

 Take any $N\ge1$ and $\alpha_k\in\Q,
k=1,\ldots,N$ with $|\alpha_1|+\cdots+|\alpha_N|>0$. It suffices
to show that   the set  $\aA_{\alpha,N}$ of all functions $Q\in
C^\ty(\overline{D},\R)$ such that the eigenvalues $\la_{j,Q}$,
$j=1,\ldots,N$ are simple and
$$\sum_{k=1}^N\alpha_k e_{k,Q}^2\not\equiv 0
$$is open and dense in $C^\ty(\overline{D},\R)$. Indeed, noting that
$$
\aA\supset\bigcap_{N\ge1,\alpha\in \Q^N}\aA_{\alpha,N}
$$we complete the proof. The fact that $\aA_{\alpha,N}$ is open
follows from the continuity of $\la_{j,Q}$ and $e_{k,Q}$ with
respect to $Q\in C^\ty(\overline{D},\R)$ at any $Q_0\in
\aA_{\alpha,N}$
 (e.g., see~\cite{KATO}).

To prove that $\aA_{\alpha,N}$ is dense, we first show that the
operator $-\Delta +Q$ satisfies the hypothesis of Theorem B in
\cite{MTEY} for any $Q\in C(\overline{D},\R) $. This implies that
any functions $Q_0,Q_1 \in C(\overline{D},\R)$ can be connected by
an analytic curve $Q_s \in C(\overline{D},\R), s\in[0,1]$ such
that the spectrum of $-\Delta +Q_s$ is simple for any $s\in
(0,1)$. In particular,~$\la_{k,Q_s}$ and $e_{k,Q_s}$ are analytic
in $s \in(0,1)$. Then we show that $\aA_{\alpha,N}$ is non-empty.
Taking any $Q_1\in \aA_{\alpha,N}$,  we see that, by
analyticity,
  also $Q_s\in \aA_{\alpha,N}$   for all $s\in[0,1]\backslash
I$, where $I\subset[0,1] $ is an at most countable set. Thus
$Q_{s_n}\rightarrow Q_0$ and $Q_{s_n}\in \aA_{\alpha,N}$ for any
$s_n\rightarrow0$ such that $s_n\in[0,1]\backslash I$.

\vspace{6pt}\textbf{Step 1.} The family $-\Delta +Q$, $Q\in
C(\overline{D},\R)$ satisfies the hypothesis of Theorem B in
\cite{MTEY}. Indeed,  the function $Q$ is  in the separable Banach
space $C (\overline{D},\R)$, and the operator $-\Delta +Q$    is
self-adjoint in $L^2(D)$ with spectrum which is discrete, of
finite multiplicity, and without finite accumulation points. Thus
it remains to show that the condition SAH2 in \cite{MTEY} is also
verified. Notice that for all $Q,P\in C(\overline{D},\R)$ and
$\e\in\R$ we have
$$
\frac{\dd }{\dd \e}(-\Delta+Q+\e P)=P.
$$Hence we have to prove that for any eigenvalue $\la$ of
$-\Delta+Q$ of multiplicity $n\ge2$ there exist two orthonormal
eigenfunctions $v_1$ and $v_2$   corresponding to~$\la$ such that
the functionals $P\rightarrow\lag P, v_1^2-v_2^2\rag$ and
$P\rightarrow\lag P, v_1 v_2 \rag$ are linearly independent.
Suppose, by contradiction, that for some eigenfunctions $v_1$ and
$v_2$ we have
$$
v_1^2(x)-v_2^2(x)-c v_1(x) v_2(x)=0 \,\,\,\text{for all $x\in D$,}
$$ where $c\in\R$ is a  constant. Thus
$$
(v_1-\frac{c}{2}v_2)^2=\frac{c^2+4}{4}v_2^2,
$$where $v_1-\frac{c}{2}v_2$ and $\frac{\sqrt{c^2+4}}{2}v_2$ are linearly independent eigenfunctions of $-\Delta+Q$ corresponding
to the eigenvalue $\la$.  Combining this with the unique
continuation theorem for the operator $-\Delta+Q$ (see \cite{JK}),
we get that $v_1-\frac{c}{2}v_2=\pm \frac{\sqrt{c^2+4}}{2}v_2 $,
which contradicts   the fact that  $v_1$ and $v_2$ are linearly
independent. Thus the functionals $\lag P, v_1^2-v_2^2\rag$ and
$\lag P, v_1 v_2 \rag$ are linearly independent. Applying Theorem
B in~\cite{MTEY}, we see that any $Q_0,Q_1 \in C(\overline{D},\R)$
can be connected by an analytic curve $Q_s \in C(\overline{D},\R),
s\in[0,1]$ such that the spectrum of $-\Delta +Q_s$ is simple for
any $s\in (0,1)$.

 \vspace{6pt}\textbf{Step 2.} To show that $\aA_{\alpha,N}$ is
 non-empty, we use the following result from the inverse spectral theory for Sturm--Liouville problems.
\begin{theorem}\label{T:Piv}Let $\{\la_{k}\}_{k=1}^N$ and
$\{\la_{k}'\}_{k=1}^N$ be two sets of positive constants such that
\begin{equation}\label{E:anhav}
\la_1<\la_1'<\la_2<\la_2'<\ldots.
 \end{equation}Then for any
$a>0$ and $n\ge1$ there is a function $W\in  L^2
([-2^na,2^na],\R)$ such that
\begin{equation}\label{E:anhav2}
\la_{k}=\la_{k,W}^{(-2^na,2^na)},\,\,\,\,\,\la_{k}'=\la_{k,W}^{(0,a)},\,\,\,\,\,k=1,\ldots,N.
\end{equation}
\end{theorem}
This theorem is a consequence of a much stronger result by
Pivovarchik (see Theorem 2.1 in \cite{VPIVO}).

Without loss of generality, we can assume that $0\in D$. Let us
choose $a>0$ and $n\ge1$ such that
\begin{equation}\label{E:Mnmx}
B':= (0,a)^m\subset D \subset  (-2^na,2^na)^m=:B.
\end{equation} By the min-max principle (e.g., see \cite{RESI}) and
(\ref{E:Mnmx}), we have
 \begin{equation}\label{E:Mnmx7}
 \la_{k,Q}^B\le \la_{k,Q}^D\le\la_{k,Q}^{B'}
 \end{equation} for any
$Q\in L^2( {B},\R)$ and $k\ge1$. Let us suppose that $Q$ is of the
form $$Q(x_1,\ldots,x_m)=P(x_1)+R(x_2)+\ldots+ R(x_m),$$ where
$c>0$ is a constant and $P,R\in L^2([-2^na,2^na])$. Then the
eigenvalues are of the form
\begin{align}
\la_{k,Q}^{B'}&=\la_{i_1,P}^{(0,a)}+\la_{i_2,R}^{(0,a)}+\ldots+\la_{i_m,R}^{(0,a)},\nonumber\\
\la_{k,Q}^{B}&=\la_{j_1,P}^{(-2^na,2^na)}+\la_{j_2,R}^{(-2^na,2^na)}+\ldots+\la_{j_m,R}^{(-2^na,2^na)}\nonumber
\end{align}
for some integers $i_p,j_p\ge1$, $p=1,\dots,m$.
 Let $\{\la_{k},\la_{k}'\}_{k=1}^N$ and $\{\tilde{ \la}_{k},\tilde{\la}_{k}'\}_{k=1}^N$
 be two sets of positive constants verifying
(\ref{E:anhav}). Applying Theorem \ref{T:Piv}, let $P,R\in  L^2
([-2^na,2^na],\R)$ be such that (\ref{E:anhav2}) holds for
$\{\la_{k},\la_{k}'\}_{k=1}^N$ and $\{\tilde{
\la}_{k},\tilde{\la}_{k}'\}_{k=1}^N$, respectively.   If
$\tilde{\la}_{2}>\la_N'$,   then
\begin{align}
\la_{k,Q}^{B'}&=\la_{k,P}^{(0,a)}+(m-1)\la_{1,R}^{(0,a)} =\la_{k}'+(m-1)\tilde{\la}_{1}',\label{E:wqwq1}\\
\la_{k,Q}^{B}&=\la_{k,P}^{(-2^na,2^na)}+(m-1)\la_{1,R}^{(-2^na,2^na)}
=\la_{k} +(m-1)\tilde{\la}_{1} \label{E:wqwq2}\end{align} for
$k=1,\ldots, N$.  We can choose $\{\la_{k},\la_{k}'\}_{k=1}^N$ and
$\{\tilde{ \la}_{k},\tilde{\la}_{k}'\}_{k=1}^N$
 such that \begin{equation}\label{E:wqwq4} \sum_{k=1}^N\alpha_k \mu_{k}\neq 0
\end{equation}for all $\mu_k\in[\la_{k,Q}^{B},\la_{k,Q}^{B'}]$. Indeed, we
deduce from (\ref{E:wqwq1}) and (\ref{E:wqwq2})
\begin{align}\label{E:wqwq3}| \sum_{k=1}^N\alpha_k \mu_{k}- \sum_{k=1}^N\alpha_k \la_{k,Q}^{B'}|&\le\sum_{k=1}^N|\alpha_k| |\mu_k-\la_{k,Q}^{B'}|
\le\sum_{k=1}^N|\alpha_k|
(\la_{k,Q}^{B'}-\la_{k,Q}^{B})\nonumber\\&\le(
 {\sup_{k\in[1,N]}}(\la_{k}'-\la_{k}) +(m-1) (\tilde{\la}_{1}'-\tilde{\la}_{1})
  )\sum_{k=1}^N|\alpha_k|.\end{align}Take
 any $\e>0$ and choose $\{\la_{k},\la_{k}'\}_{k=1}^N$ and $\{\tilde{ \la}_{1},\tilde{\la}_{1}'\} $ such that
\begin{equation}\label{E:poty}
(
 {\sup_{k\in[1,N]}}(\la_{k}'-\la_{k}) +(m-1) (\tilde{\la}_{1}'-\tilde{\la}_{1})
  )\sum_{k=1}^N|\alpha_k|<\e.
\end{equation}On the other hand, we can choose $\{\la_{k}'\}_{k=1}^N$ and
$\{\tilde{\la}_{k}'\}_{k=1}^N$ such that
$$
|\sum_{k=1}^N\alpha_k \la_{k,Q}^{B'}|>  \e .
$$Combining this with (\ref{E:wqwq3}) and (\ref{E:poty}), we arrive at
(\ref{E:wqwq4}).
 Thus (\ref{E:Mnmx7}) implies that
 \begin{equation}\label{E:anhav3} \sum_{k=1}^N\alpha_k
\la_{k,Q}^D\neq 0.
\end{equation}Without loss of generality, we can assume that $Q\in
C^\ty(\overline{D},\R)$.
 Take any
$\tilde{Q}\in  C^\ty(\overline{D},\R)\backslash\aA_{\alpha,N}$ (if
$ C^\ty(\overline{D},\R)=\aA_{\alpha,N}$ then the proof is
completed). By Step 2, the functions $Q$ and $\tilde{Q}$ can be
connected by an analytic curve $\tilde{Q}_s \in
C(\overline{D},\R), s\in(0,1)$ such that the spectrum of $-\Delta
+\tilde{Q}_s$ is simple for any $s\in (0,1)$. We deduce from
(\ref{E:anhav3}) that the analytic function $ \sum_{k=1}^N\alpha_k
\la_{k,\tilde{Q}_s}^D $  is non-constant on $[0,1]$. This implies
that
\begin{equation}\label{E:ioew} 0\neq \frac{\dd}{\dd s}\sum_{k=1}^N\alpha_k \la_{k,\tilde{Q}_s}^D=\lag\frac{\dd \tilde{Q}_s}{\dd s},\sum_{k=1}^N\alpha_k
(e_{k,\tilde{Q}_s}^D)^2\rag
\end{equation}for all $s\in[0,1]\backslash\tilde{I}$, where $\tilde{I}\subset[0,1] $ is an at most countable set  (cf. Step 1 of the proof of Theorem 3.1).
Indeed, taking the derivative of  the identity
$$(-\Delta+\tilde{Q}_s  -\la_{k,  \tilde{Q}_s }^D)e_{k,  \tilde{Q}_s }^D=0$$
with respect to $s $, we get
$$
(-\Delta+\tilde{Q}_s -\la_{k, \tilde{Q}_s}^D)\frac{\dd
e_{k,\tilde{Q}_s }^D }{\dd s}  + (\frac{\dd \tilde{Q}_s}{\dd
s}-\frac{\dd \la_{k,\tilde{Q}_s}^D }{\dd s}) e_{k,\tilde{Q}_s
}^D=0.
$$Taking the scalar product of this identity with $e_{k,\tilde{Q}_s}^D$, we
obtain
\begin{equation}\label{E:ppa}
\frac{\dd \la_{k,\tilde{Q}_s }^D }{\dd s} = \langle \frac{\dd
\tilde{Q}_s}{\dd s}, ( e_{k,\tilde{Q}_s}^D) ^2  \rangle,
\end{equation}
which implies (\ref{E:ioew}).
 Finally, (\ref{E:ioew}) shows
that
$$ \sum_{k=1}^N\alpha_k
(e_{k,\tilde{Q}_s}^D)^2 \neq 0
$$for all $s\in[0,1]\backslash\tilde{I}$ and $\tilde{Q}_s\in
\aA_{\alpha,N}.$ Thus $\aA_{\alpha,N}$ is non-empty.
\end{proof}


\addcontentsline{toc}{section}{References}
\begin{thebibliography}{10}

\bibitem{AC}
A.~Agrachev and T.~Chambrion.
\newblock {An estimation of the controllability time for single-input systems
  on compact Lie groups}.
\newblock {\em J. ESAIM Control Optim. Calc. Var.}, 12(3):409--441, 2006.

\bibitem{ALBERT}
J.~H. Albert.
\newblock {Genericity of simple eigenvalues for elliptic PDE's}.
\newblock {\em Proc. Amer. Math. Soc.}, 48:413--418, 1975.

\bibitem{ALAL}
F.~Albertini and D.~D'Alessandro.
\newblock {Notions of controllability for bilinear multilevel quantum systems}.
\newblock {\em IEEE Transactions on Automatic Control}, 48(8):1399--1403, 2003.

\bibitem{ALT}
C.~Altafini.
\newblock {Controllability of quantum mechanical systems by root space
  decomposition of $su(n)$}.
\newblock {\em J. of Math. Phys.}, 43(5):2051--2062, 2002.

\bibitem{BMS}
J.~M. Ball, J.~E. Marsden, and M.~Slemrod.
\newblock {Controllability for distributed bilinear systems}.
\newblock {\em SIAM J. Control Optim.}, 20(4):575--597, 1982.

\bibitem{BP}
L.~Baudouin and J.-P. Puel.
\newblock {Uniqueness and stability in an inverse problem for the Schr\"odinger
  equation}.
\newblock {\em Inverse Problems}, 18(6):1537--1554, 2001.

\bibitem{BCH}
K.~Beauchard.
\newblock {Local controllability of a 1-D Schr\"odinger equation}.
\newblock {\em J. Math. Pures et Appl.}, 84(7):851--956, 2005.

\bibitem{BPE}
K.~Beauchard.
\newblock {Personal communication}.
\newblock 2009.

\bibitem{BeCo}
K.~Beauchard and J.-M. Coron.
\newblock {Controllability of a quantum particle in a moving potential well}.
\newblock {\em J. Funct. Anal.}, 232(2):328--389, 2006.

\bibitem{BCMR}
K.~Beauchard, J.-M. Coron, M.~Mirrahimi, and P.~Rouchon.
\newblock {Implicit Lyapunov control of finite dimensional Schr\"odinger
  equations}.
\newblock {\em Systems and Control Letters}, 56(5):388--395, 2007.

\bibitem{BM}
K.~Beauchard and M.~Mirrahimi.
\newblock {Approximate stabilization of a quantum particle in a 1D infinite
  square potential well}.
\newblock {\em Submitted}, 2007.

\bibitem{Bou1}
J.~Bourgain.
\newblock {Periodic nonlinear Schr\"odinger equation and invariant measures}.
\newblock {\em Comm. Math. Phys.}, 166(1):1--26, 1994.

\bibitem{CW}
T.~Cazenave.
\newblock {Semilinear Schr\"odinger equations}.
\newblock {\em Courant Lecture Notes in Mathematics, AMS}, 10, 2003.

\bibitem{CH}
T.~Chambrion, P.~Mason, M.~Sigalotti, and U.~Boscain.
\newblock {Controllability of the discrete-spectrum Schr\"odinger equation
  driven by an external field}.
\newblock {\em Annales de l'IHP, Non Linear Analysis}, 26(1):329--349, 2009.

\bibitem{DOd}
A.~Debussche and C.~Odasso.
\newblock {{Ergodicity for a weakly damped stochastic nonlinear Schr\"odinger
  equations}}.
\newblock {\em J. Evol. Eq.}, 3(5):317--356, 2005.

\bibitem{DGL}
B.~Dehman, P.~G\'erard, and G.~Lebeau.
\newblock {Stabilization and control for the nonlinear Schr\"odinger equation
  on a compact surface}.
\newblock {\em Math. Z.}, 254(4):729--749, 2006.

\bibitem{PERV}
S.~Ervedoza and J.-P. Puel.
\newblock {Approximate controllability for a system of Schrödinger equations
  modeling a single trapped ion}.
\newblock {\em Annales de l'IHP, Non Linear Analysis}, to appear.

\bibitem{JK}
D.~Jerison and C.~E. Kenig.
\newblock {Unique continuation and absence of positive eigenvalues for
  Schr\"{o}dinger operators (with an appendix by E. M. Stein)}.
\newblock {\em Ann. of Math.}, 121(3):463--494, 1985.

\bibitem{KATO}
T.~Kato.
\newblock {Perturbation Theory for Linear Operators}.
\newblock {\em Berlin, Springer}, 1995.

\bibitem{YK}
Y.~Kifer.
\newblock {Ergodic Theory of Random Transformations}.
\newblock {\em Birkhauser}, 1986.

\bibitem{KS2001}
S.~Kuksin and A.~Shirikyan.
\newblock {Ergodicity for the randomly forced 2D Navier--Stokes equations}.
\newblock {\em Math. Phys. Anal. Geom.}, 4(2):147--195, 2001.

\bibitem{KUKSHIR}
S.~Kuksin and A.~Shirikyan.
\newblock Randomly forced {CGL} equation: stationary measures and the inviscid
  limit.
\newblock {\em J. Phys. A: Math. Gen.}, 37(12):3805--3822, 2004.

\bibitem{L}
G.~Lebeau.
\newblock {Contr\^{o}le de l'\'equation de Schr\"odinger}.
\newblock {\em J. Math. Pures Appl.}, 71(3):267--291, 1992.

\bibitem{MZ}
E.~Machtyngier and E.~Zuazua.
\newblock {Stabilization of the Schr\"odinger equation}.
\newblock {\em Portugaliae Matematica}, 51(2):243--256, 1994.

\bibitem{Mir}
M.~Mirrahimi.
\newblock {Lyapunov control of a particle in a finite quantum potential well}.
\newblock {\em IEEE Conf. on Decision and Control, San Diego}, 2006.

\bibitem{N}
V.~Nersesyan.
\newblock {{Exponential} {mixing} {for} {finite-dimensional} {approximations}
  {of} {the} {Schr\"odinger} {equation with multiplicative noise}}.
\newblock {\em Submitted}, 2008.

\bibitem{VN}
V.~Nersesyan.
\newblock {Growth of Sobolev norms and controllability of Schr\"odinger
  equation}.
\newblock {\em Comm. Math. Phys.}, to appear.

\bibitem{OKS}
B.~{\O}ksendal.
\newblock {Stochastic Differential Equations}.
\newblock {\em Springer--Verlag}, 2003.

\bibitem{VPIVO}
V.~N. Pivovarchik.
\newblock {An inverse Sturm--Liouville problem by three spectra}.
\newblock {\em Integr. Equ. Oper. Theory}, 34(2):234--243, 1999.

\bibitem{PISI}
Y.~Privat and M.~Sigalotti.
\newblock {The squares of Laplacian--Dirichlet eigenfunctions are generically
  linearly independent}.
\newblock {\em Preprint}, 2008.

\bibitem{RSDR}
V.~Ramakrishna, M.~Salapaka, M.~Dahleh, H.~Rabitz, and A.~Pierce.
\newblock {Controllability of molecular systems}.
\newblock {\em Phys. Rev. A}, 51(2):960--966, 1995.

\bibitem{RESI}
M.~Reed and B.~Simon.
\newblock {Methods of Modern Mathematical Physics. Vol. 4: Analysis of
  Operators}.
\newblock {\em Academic Press, New York}, 1978.

\bibitem{MTEY}
M.~Teytel.
\newblock {How rare are multiple eigenvalues?}
\newblock {\em Comm. Pure Appl. Math.}, 52(8):917--934, 1999.

\bibitem{TR}
G.~Turinici and H.~Rabitz.
\newblock {Quantum wavefunction controllability}.
\newblock {\em Chem. Phys.}, 267(1):1--9, 2001.

\bibitem{NTZ1}
N.~Tzvetkov.
\newblock {Invariant measures for the nonlinear Schr\"odinger equation on the
  disc}.
\newblock {\em Dynamics of PDE}, 3(2):111--160, 2006.

\bibitem{Z}
E.~Zuazua.
\newblock {Remarks on the controllability of the Schr\"odinger equation}.
\newblock {\em CRM Proc. Lecture Notes}, 33:193--211, 2003.

\end{thebibliography}
\end{document}